\documentclass{ws-procs9x6}

\begin{document}

\title{MODERN IVES-STILWELL EXPERIMENTS AT STORAGE RINGS:
LARGE BOOSTS MEET HIGH PRECISION}

\author{G.\ GWINNER}

\address{Department of Physics and Astronomy, University of Manitoba\\
Winnipeg, R3T 2N2, Canada\\
E-mail: gwinner@physics.umanitoba.ca}

\author{B.\ BOTERMANN, C.\ GEPPERT, G.\ HUBER, S.\ KARPUK, A.\ KRIEGER,\\
W.\ N\"ORTERSH\"AUSER, and C.\ NOVOTNY}

\address{Johannes Gutenberg-Universit\"at Mainz,
55128 Mainz, Germany}

\author{T.\ K\"UHL, R. S\'ANCHEZ, and T.\ ST\"OHLKER}

\address{GSI Helmholtzzentrum f\"ur Schwerionenforschung,
64291 Darmstadt, Germany}

\author{D.\ BING, D.\ SCHWALM, and A.\ WOLF}

\address{Max-Planck Institut f\"ur Kernphysik,
69117 Heidelberg, Germany}

\author{T.W.\ H\"ANSCH, S.\ REINHARDT, and G.\ SAATHOFF}

\address{Max-Planck Institut f\"ur Quantenoptik,
85748 Garching, Germany}

\begin{abstract}
We give a brief overview of time dilation tests using high-resolution laser spectroscopy at heavy-ion storage rings. We reflect on the various methods used to eliminate the first-order Doppler effect and on the pitfalls encountered, and comment on possible extensions at future facilities providing relativistic heavy ion beams at $\gamma \gg 1$.
\end{abstract}

\bodymatter

\section{Introduction}

Time dilation measurements via precise optical spectroscopy have played a long and distinguished role in testing the predictions of Special Relativity (SR), and have recently also contributed to constraining coefficients in the Lorentz- and CPT-violating Standard-Model Extension (SME)~\cite{reinhardt07,lane05}. The first reported observation of time dilation in SR in 1938, by Ives and Stilwell, was based on measuring the relativistic Doppler effect in fast hydrogen atoms~\cite{is38}. In the decades after this landmark experiment, other methods tested time dilation with increasing sensitivity, such as lifetime measurements of elementary particles accelerated to relativistic energies, transverse Doppler effect measurements with rotating M\"{o}ssbauer setups, and comparisons between ground-based and rocket-mounted clocks (for a review and references see, e.g., Ref.\ \refcite{gwinner05}). Two major breakthroughs allowed optical spectroscopy to re-enter the field, and to dominate it since: tunable, single-mode lasers in the 1970s and heavy-ion storage rings, providing electron-cooled ion beams of exquisite energy definition and transverse emittance, in the 1980s.
This has led to the measurement of absolute optical transition frequencies at the $10^{-10}$ level on near-relativistic beams~\cite{reinhardt07}, possibly the most accurate measurement ever performed with accelerated particle beams.

\section{Relativistic Doppler shift measurements}

An atom or ion is moving with velocity $\vec{\beta}$ in the laboratory. A transition between two levels in this atom can be driven in its rest frame by radiation of frequency $\nu_0$. A light source in the laboratory frame, according to the relativistic Doppler effect, needs to be tuned to the frequency $\nu_{\rm lab}$ such that
\begin{equation}
\nu_0 = \gamma \left(1-\beta \cos{\theta}  \right) \nu_{\rm lab},
\label{gg:eq1}
\end{equation}
where $\beta$ is the relative velocity between the atom and the laboratory frame, and $\gamma = 1/\sqrt{1-\beta^2}$; $\theta$ is the angle between $\vec{\beta}$ and the wavevector of the radiation exciting the atom, as measured in the laboratory frame. Testing time dilation, i.e., $\gamma$, requires the precise determination of 
$\nu_0$, $\nu_{\rm lab}$, $\beta$, and $\theta$. In reality, it is generally not possible to determine the velocity of an accelerated particle beam with the required precision. Therefore, experiments need to remove the first-order dependence on $\beta$. At $\theta = \pi/2$ one could observe a purely transverse Doppler effect, but due to the sensitivity to angular misalignment, this turns out not to be a workable scheme.
Geometries with $\theta = (0, \pi)$, while robust against misalignments, appear to suffer from the presence of the full first-order Doppler effect. Nevertheless, atomic spectroscopy provides several well-established techniques to eliminate the first-order Doppler effect, turning a collinear setup into the method of choice.

\begin{figure}
\begin{center}
\psfig{file=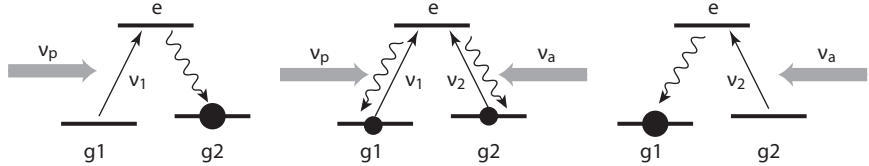,width=\textwidth}
\end{center}
\caption{$\Lambda$-spectroscopy: driving a given velocity class only on transition 1 (2) pumps the atoms quickly into the `dark' state g2 (g1) as shown on the left (right). When both legs are excited, atoms get continually pumped back and forth and strong fluorescence is produced (center).}
\label{gg:fig1}
\end{figure}

Commercially available single-mode lasers have much higher monochromaticity ($\Delta\nu/\nu < 10^{-9}$) than accelerated particle beams ($\Delta \beta/\beta \approx 10^{-5}$-$10^{-6}$). In a scenario with two lasers, one (at a fixed frequency) can be used to `mark' a narrow velocity class within the particle beam's velocity distribution such that a second, frequency-tunable, laser performs spectroscopy only on that specific class of atoms. 

The first generation of storage ring measurements (1990-94) used Li$^+$ stored at $\beta = 0.064$ in the TSR storage ring~\cite{grieser94}. It used a Doppler-free technique known as $\Lambda$-spectroscopy, which is illustrated in Fig.\ 1.
Two lasers with laboratory frequencies $\nu_a$ and $\nu_p$ counter- and co-propagate with the ions and excite both legs of the $\Lambda$ at the frequencies $\nu_1$ and $\nu_2$ in the ions' rest frame. Only if both lasers are exciting the {\em same} velocity class do atoms get pumped back and forth, producing significant fluorescence.
Within SR, the resonance conditions
$\nu_1=\gamma (1-\beta)\nu_p$ and $\nu_2=\gamma (1+\beta)\nu_a$ give the relation 
$\nu_1 \nu_2 = \nu_a \nu_p$. Deviations from relativity can be parametrized in a model-independent way by a velocity-dependent term $\epsilon(\beta^2)$ as
\begin{equation}
\sqrt{\frac{\nu_a \nu_p}{\nu_1 \nu_2}} =  1 + \epsilon(\beta^2),
\label{gg:eq2}
\end{equation}
which can be expanded, for $\beta \ll 1$, as $\epsilon(\beta^2) \approx \hat{\alpha} \beta^2 + \hat{\alpha}_2 \beta^4$. The TSR-1 experiment yielded a new, best time dilation limit of $|\hat{\alpha}| < 7 \times 10^{-7}$~\cite{grieser94}. The experimental limitation was an unexpectedly broad lineshape, with a width of 54 MHz, vs. 7.6 MHz expected from the natural linewidth of the transition. The second-generation TSR-2 experiment (2000-06) revealed the problem: the `marking' of a velocity class by one laser and the `sampling' of it by the other are generally not concurrent in $\Lambda$-spectroscopy and velocity-changing collisions lead to broadening. 

The TSR-2 experiment solved this problem by employing a different Doppler-free technique: saturation spectroscopy~\cite{saathoff03,reinhardt07}. Its signature feature, the Lamb dip, only occurs when both lasers {\em simultaneously} interact with the same individual ion. Linewidths approaching the natural linewidth were observed, and a much improved limit was obtained: $|\hat{\alpha}| < 9 \times 10^{-8}$~\cite{reinhardt07}. Ultimately, the method is limited by its reliance on saturation, i.e., the need to scatter many photons. At the intensities required for saturation, the lasers exert a rather strong force on the ions, distorting the velocity distribution significantly. In saturation spectroscopy, we observe a small {\em drop} in fluorescence (the Lamb dip) on top of a large Doppler-broadened background. In $\Lambda $-spectroscopy, we detect a narrow fluorescence {\em peak}, while the rest of the velocity distribution remains dark.

With the spectroscopy reaching its limitations, higher ion velocities are the obvious avenue for further improvements. At the storage ring ESR, the same experiment can be carried out at $\beta = 0.338$, where the Doppler shifts are extraordinary~\cite{novotny09,botermann13}: For $\lambda_{1,2} = 548$ nm, the laser frequencies in the laboratory frame are $\lambda_p = 386$ nm and $\lambda_a = 772$ nm. Saturation spectroscopy turned out not to be viable due to the required intensity for the UV light. $\Lambda$-spectroscopy proved robust; while the excess broadening was comparable to the TSR-1 situation ($\approx 50$ MHz), the much larger $\beta$ more than compensates and tighter limits result: an upper limit for $|\hat{\alpha}|$ in the lower $10^{-8}$ range seems in reach~\cite{botermann13}. Alternatively, TSR-2 and ESR will be able to set combined limits of $|\hat{\alpha}| < 9 \times 10^{-8}$ and $|\hat{\alpha_2}| < 10^{-6}$.

\section{Prospects for future work}

Ultimately, it would be intriguing to implement high-resolution spectroscopy at facilities that provide highly relativistic ion beams, like the upcoming High-Energy Storage Ring (HESR) at FAIR. If the rest-frame frequencies $\nu_{1,2}$ are nearly identical and in the visible region, at high boosts
$\nu_{p}$ will eventually be beyond the near-UV region. A possible solution would be an {\em asymmetric} $\Lambda$, such that standard lasers can drive it in a low-$\beta$ measurement (e.g., one in the visible and one in the near/mid-infrared). In the high-boost version, the co-propagating laser will drive the short leg (which will be blue-shifted toward the visible) and the counter-propagating laser will excite the long leg (which will be red-shifted toward the infrared). The challenge will be to identify a suitable transition in a highly charged ion.


\begin{thebibliography}{x}

\bibitem{reinhardt07} 
S.\ Reinhardt {\em et al.}, Nat.\ Phys.\ {\bf 3}, 861 (2007).

\bibitem{lane05}
C.D.\ Lane, Phys.\ Rev.\ D {\bf 72}, 016005 (2005).

\bibitem{is38}
H.E.\ Ives and G.R.\ Stilwell, J.\ Opt.\ Soc.\ Am. {\bf 28}, 215 (1938).

\bibitem{gwinner05}
G.\ Gwinner, Mod.\ Phys.\ Lett.\ A {\bf 20}, 791 (2005).

\bibitem{grieser94}
R.\ Grieser {\em et al.}, Appl.\ Phys.\ B {\bf 59}, 127 (1994).

\bibitem{saathoff03}
G.\ Saathoff {\em et al.}, Phys.\ Rev.\ Lett.\ {\bf 91}, 190403 (2003).

\bibitem{novotny09}
Ch.\ Novotny {\em et al.}, Phys.\ Rev.\ A {\bf 80}, 022107 (2009).

\bibitem{botermann13}
B.\ Botermann {\em et al.}, in preparation.

\end{thebibliography}
\end{document}